\documentclass{article}
\usepackage{amsmath, amssymb, amsfonts}
\title{THE ZERO POINT ENERGY AND GRAVITATION}
\author{Hristu Culetu, \\Ovidius University, Dept.of Physics, \\B-dul Mamaia 124, 8700 Constanta, Romania, \\e-mail : hculetu@yahoo.com}

\begin{document}
\numberwithin{equation}{section}
\pagenumbering{arabic}
\maketitle
\newcommand{\fv}{\boldsymbol{f}}
\newcommand{\tv}{\boldsymbol{t}}
\newcommand{\gv}{\boldsymbol{g}}
\newcommand{\OV}{\boldsymbol{O}}
\newcommand{\wv}{\boldsymbol{w}}
\newcommand{\WV}{\boldsymbol{W}}
\newcommand{\NV}{\boldsymbol{N}}
\newcommand{\hv}{\boldsymbol{h}}
\newcommand{\yv}{\boldsymbol{y}}
\newcommand{\RE}{\textrm{Re}}
\newcommand{\IM}{\textrm{Im}}
\newcommand{\rot}{\textrm{rot}}
\newcommand{\dv}{\boldsymbol{d}}
\newcommand{\grad}{\textrm{grad}}
\newcommand{\Tr}{\textrm{Tr}}
\newcommand{\ua}{\uparrow}
\newcommand{\da}{\downarrow}
\newcommand{\ct}{\textrm{const}}
\newcommand{\xv}{\boldsymbol{x}}
\newcommand{\mv}{\boldsymbol{m}}
\newcommand{\rv}{\boldsymbol{r}}
\newcommand{\kv}{\boldsymbol{k}}
\newcommand{\VE}{\boldsymbol{V}}
\newcommand{\sv}{\boldsymbol{s}}
\newcommand{\RV}{\boldsymbol{R}}
\newcommand{\pv}{\boldsymbol{p}}
\newcommand{\PV}{\boldsymbol{P}}
\newcommand{\EV}{\boldsymbol{E}}
\newcommand{\DV}{\boldsymbol{D}}
\newcommand{\BV}{\boldsymbol{B}}
\newcommand{\HV}{\boldsymbol{H}}
\newcommand{\MV}{\boldsymbol{M}}
\newcommand{\be}{\begin{equation}}
\newcommand{\ee}{\end{equation}}
\newcommand{\ba}{\begin{eqnarray}}
\newcommand{\ea}{\end{eqnarray}}
\newcommand{\bq}{\begin{eqnarray*}}
\newcommand{\eq}{\end{eqnarray*}}
\newcommand{\pa}{\partial}
\newcommand{\f}{\frac}
\newcommand{\FV}{\boldsymbol{F}}
\newcommand{\ve}{\boldsymbol{v}}
\newcommand{\AV}{\boldsymbol{A}}
\newcommand{\jv}{\boldsymbol{j}}
\newcommand{\LV}{\boldsymbol{L}}
\newcommand{\SV}{\boldsymbol{S}}
\newcommand{\av}{\boldsymbol{a}}
\newcommand{\qv}{\boldsymbol{q}}
\newcommand{\QV}{\boldsymbol{Q}}
\newcommand{\ev}{\boldsymbol{e}}
\newcommand{\uv}{\boldsymbol{u}}
\newcommand{\KV}{\boldsymbol{K}}
\newcommand{\ro}{\boldsymbol{\rho}}
\newcommand{\si}{\boldsymbol{\sigma}}
\newcommand{\thv}{\boldsymbol{\theta}}
\newcommand{\bv}{\boldsymbol{b}}
\newcommand{\JV}{\boldsymbol{J}}
\newcommand{\nv}{\boldsymbol{n}}
\newcommand{\lv}{\boldsymbol{l}}
\newcommand{\om}{\boldsymbol{\omega}}
\newcommand{\Om}{\boldsymbol{\Omega}}
\newcommand{\Piv}{\boldsymbol{\Pi}}
\newcommand{\UV}{\boldsymbol{U}}
\newcommand{\iv}{\boldsymbol{i}}
\newcommand{\nuv}{\boldsymbol{\nu}}
\newcommand{\muv}{\boldsymbol{\mu}}
\newcommand{\lm}{\boldsymbol{\lambda}}
\newcommand{\Lm}{\boldsymbol{\Lambda}}
\newcommand{\opsi}{\overline{\psi}}
\renewcommand{\tan}{\textrm{tg}}
\renewcommand{\cot}{\textrm{ctg}}
\renewcommand{\sinh}{\textrm{sh}}
\renewcommand{\cosh}{\textrm{ch}}
\renewcommand{\tanh}{\textrm{th}}
\renewcommand{\coth}{\textrm{cth}}

 A possible connection between the energy W of the vacuum fluctuations of quantum fields and gravity in ``empty space'' is conjectured in this paper using a natural cutoff of high momenta with the help of the gravitational radius of the vacuum region considered. We found that below some ``critical'' length $L = 1 mm$ the pressure $\sigma$ is one third of the energy density $\epsilon$, as for dark matter, but above $1~ mm$ the equation of state is $\sigma = -\epsilon$ (dark energy). 
 In the case of a massive field, W does not depend on the mass of the field for $L << 1 mm$ but for $L >> 1 mm$ it does not depend on $\hbar$. In addition, when the Newton constant tends to zero, W becomes infinite. The energy density is also a function of the volume V of the vacuum region taken into account.
 \\Keywords : UV cutoff, Schwarzschild radius, dark energy, holographic principle, quantum gravity.

 \section{INTRODUCTION}
 
 The zero point energy was firstly introduced in physics by Max Planck.
The average energy $\overline{E}$ of a harmonic oscillator at temperature T would be given by \cite {DS}
\begin{equation}
\overline{E} = \frac{\hbar\omega}{2} + \frac{\hbar\omega}{\exp(\frac{\hbar\omega}{k_{B}T})-1}
\label{1.1}
\end{equation}
 One means the oscillator has an energy $\hbar\omega/2$ even at $T = 0$, where T is the temperature of the black body radiation. $\omega$ is the frequency of the oscillator and $k_{B}$ and $\hbar$ - the Boltzmann and the Planck constant, respectively.
 \\ A very important effect of the zero point energy consists in maintaining helium in the liquid state under its vapour pressure at $T = 0$. As D. Sciama \cite {DS} pointed out,``the zero point motion of the atoms keeps them sufficiently far apart on the average so that the attractive forces between them are too weak to cause solidification (even close to absolute zero helium is hot enough to be liquid'').
 \\The boundary conditions associated with a physical system narrow the range of normal modes which contribute to the ground state of the system and so to the zero point energy. For instance, some of normal modes of a massless field (say, the electromagnetic field) are excluded by boundary conditions (the experimentally proven Casimir effect). The modes whose wavelength exceeds the distance between the parallel conductors cannot propagate inside the system.
 \\The zero point energy associated with the quantum fluctuations of a massless or massive field is infinite due to the lack of a natural cutoff of the high frequences. The common procedure is to use an exponential cutoff \cite {LF} to render finite the vacuum expectation values $<0|T_{\mu\nu}|0>$ of the stress tensor of the quantum field. 
 \\In 1973, J.D.Bekenstein conjectured that for a system of energy M localized in a region of linear dimension R, its entropy is bounded from above
\begin{equation}
S < \frac{2\pi c k_{B}MR}{\hbar}
\label{1.2}
\end{equation}
 For weak gravity, the linear dimensions of the system are much larger than its Schwarzschild radius $r_{g} = 2GM/c^{2}$. Therefore,
\begin{equation}
S < \frac{c^{3} k_{B}}{4 G \hbar} A ,
\label{1.3}
\end{equation}
where $A = 4 \pi R^{2}$ is the surface area of the system \cite {HH}. Eq. (1.3) is in accordance with the Holographic Principle \cite {RB} which states that the entropy in a spatial volume V with surface A as its boundary cannot exceed $A/4$.
 \\As Hong and Hsu have noticed, the energy of a system of size R must have an upper bound not to collapse into a black hole. In other words, the linear dimension of the physical system has to be longer than its gravitational radius.
 \\A. Aste \cite {AA} reached similar conclusions by imposing a cutoff on the maximum energy of the field modes of the order of the Planck energy. The maximum energy $W_{max}$ of a state in Fock space should be $M_{bh}\left(R\right) c^{2}$, the energy of a black hole with radius R. Therefore \cite {AA}
\begin{equation}
W_{max} \approx \frac{c^{4}}{2 G} R .
\label{1.4}
\end{equation}
\\Recently, T.Padmanabhan \cite{TP} has observed that we have to treat the energy fluctuations as the physical quantity ``detected'' by gravity when quantum effects are taken into account. Quantum theory taught us that the energy density (even the vacuum state concept) depends on the scale where it is probed.
 \\On the same line as above, we try in this paper to render finite the zero point energy of quantum fields by removing the modes of short wavelengths by means of a gravitational cutoff. $\lambda_{min}$ must exceed the gravitational radius associated with the energy of the system taken into consideration. In Sec.2 we compute the finite energy W of a vacuum region of volume V, linear dimension $L = V^{1/3}$ and temporal dimension L/c. We find that $W \propto V^{1/5}$. Note that the system has no physical boundaries.
 \\In Sec.3, W is calculated for the quantum fluctuations of a massive scalar field with m its mass, taken here as the proton mass. We found that W does not depend on m for $L << 1 mm$. In addition, the pressure $\sigma$ is one third of the vacuum energy density $\epsilon$. 
 \\The expression for W is similar with that corresponding to the massless field. For $L >> 1 mm$, W is independent of the Planck constant in the approximationn used. In both cases, $\epsilon$ is not constant but a function of the independent variable V. We show that for the macroscopic case, when Einstein's equations are important, we have $\sigma + \epsilon = 0$, the well - known property which leads to a Lorentz - invariant vacuum with the appearance of a cosmological constant.
 \\We conclude with a summary of our results and a brief discussion of open questions.
 \\We use from now on geometrical units $c = G = \hbar = k_{B} = 1$. 
 
 \section{THE MASSLESS SCALAR FIELD}
 
 Let us take a vacuum region of volume V, with linear dimension L, temporal dimension L and with no physical boundaries. Let us suppose that the energy of the vacuum state of the quantized massless field in the volume V is W, a finite value.
 \\Let $T_{\mu\nu}$ be the stress tensor of the field inside V. It is well - known that the naive vacuum expectation value $<0|T_{\mu\nu}|0>$ is divergent because each field mode of angular frequency $\omega$ contributes a zero point energy $(1/2) \hbar \omega$ to the vacuum \cite {WS}. It is the renormalized expectation value $<0|T_{\mu\nu}|0>_{ren}$ from which the divergences could be removed in an invariant way. Therefore, the energy density is given by
\begin{equation}
\epsilon = \int _{0}^{\omega_{max}} u(\omega) d\omega = \frac{1}{4 \pi^{2}} \int _{0}^{\omega_{max}} \omega^{3} d\omega
\label{2.1}
\end{equation}
where $u\left(\omega\right)$ is the spectral density. The cutoff frequency $\omega_{max}$ will be determined in the following manner. As the energy inside the volume V has a finite value W, the linear dimensions of the system should be more than the gravitational radius $r_{g}$ associated with W, otherwise the system will collapse into a black hole
\begin{equation}
L \geq 2 W .
\label{2.2}
\end{equation}
Therefore
\begin{equation}
\omega_{max} = \frac{2 \pi}{\lambda_{min}} = \frac{\pi}{W} ,
\label{2.3}
\end{equation}
because the shortest wavelength of the field modes cannot exceed the Schwarzschild radius of the system. One obtains from (2.1) \footnote{Keeping in mind that W is not proportional to V, we should have used $\epsilon = dW/dV$ but the difference does not change the order of magnitude of the energy density} 

\begin{equation}
\epsilon = \frac{W}{V} = \frac{\pi^{2}}{16 W^{4}} ,
\label{2.4}
\end{equation}

whence
\begin{equation}
W\left(V\right) = \left(\frac{\pi^{2}}{16}\right)^{1/5} V^{1/5}.
\label{2.5}
\end{equation}
(Keeping track of all fundamental constants, we have $W^{5} = \pi^{2} \hbar c^{17} V/16 G^{4}$). Written in a different form, Eq. (2.5) appears as 
\begin{equation}
\left(\frac{W}{W_{P}}\right)^{5} = \frac{\pi^{2}}{16} \left(\frac{L}{L_{P}}\right)^{3},
\label{2.6}
\end{equation}
where $W_{P}$ and $L_{P}$ are Planck's energy and length, respectively.
 \\We see that the zero point energy of the massless scalar field depends only on the volume considered and fundamental constants. In addition, when G tends to zero, the energy becomes infinite. In other words, gravitation acts as a natural cutoff. For instance, the vacuum energy contained in a cube with $L = 100~ cm$ is $W \approx 10^{37}$ erg and $\epsilon \approx 10^{31} erg/cm^{3}$ for a time interval shorter than $L \approx 10^{-8}s$. The corresponding Schwarzschild radius is $r_{g} \approx 10^{-12} cm$ which leads to $\omega_{max} \approx 10^{23}s^{-1}$. It means the shortest wavelength\footnote{We should have taken $\omega_{min} = 2 \pi/L$ in Eq.(2.1) but the result is changed in a negligible manner} of the normal modes is of the order of the nuclear radius when the physical system has a volume of $1 m^{3}$.
With, instead, $L = 10^{-8}$ cm, we have $W \approx 10^{31}$ erg,~$\epsilon \approx 10^{55} erg/cm^{3}$ and $r_{g} \approx 10^{-18} cm$. Therefore,~$\omega_{max} \approx 10^{27}s^{-1}$. We observe that, although W increases with V, $\epsilon$ decreases, being very large for microscopic values of L. 
 \\It is worth to note that the above values are valid only for a time less than $L \approx 10^{-18}s$, otherwise we leave V in temporal direction. 
 \\It is an easy task to check that we always obtain from (2.6) that any $L \geq r_{g}$ leads to $L \geq L_{P}$, as it should be. It is a confirmation that the formula (2.6) works for any $L \geq  L_{P}$ and, in principle, beyond it. On the contrary, we will see that the situation is different for the massive field case.
 
\section{THE MASSIVE SCALAR FIELD}

 Let us consider now the energy of the vacuum fluctuations of a quantum scalar field of mass m to be the proton mass, $10^{-24}$ g. As is well known \cite {LF} \cite{YZ}, the relativistic energy density of the zero point fluctuations of a massive scalar field is given by
\begin{equation}
\epsilon = \frac{1}{4 \pi^{2}} \int_{0}^{\infty} p^{2} \sqrt{p^{2}+m^{2}}  dp
\label{3.1}
\end{equation}
where p is the momentum of the particle of the field. The corresponding pressure is 
\begin{equation}
\sigma = \frac{1}{3} \frac{1}{4 \pi^{2}} \int_{0}^{\infty} \frac{p^{4}}{\sqrt{p^{2}+m^{2}}} dp ,
\label{3.2}
\end{equation}
$\epsilon$ and $\sigma$ are, of course, the vacuum expectation values $<0| T^{0}_{0} |0>$ and $<0| T^{1}_{1} |0> = <0| T^{2}_{2} |0> = <0| T^{3}_{3} |0>$, respectively, the formal divergent quantities.
 \\We try in this paper to render $T^{\nu}_{\mu}$ finite with the help of a UV cutoff, using gravity. We therefore replace the expressions (3.1) and (3.2) by
\begin{equation}
\epsilon = \frac{1}{4 \pi^{2}} \int_{0}^{1/r_{g}} p^{2} \sqrt{p^{2}+m^{2}}dp
\label{3.3}
\end{equation}

and
\begin{equation}
\sigma = \frac{1}{12 \pi^{2}} \int_{0}^{1/r_{g}} \frac{p^{4}}{\sqrt{p^{2}+m^{2}}} dp
\label{3.4}
\end{equation}
where $p_{max} = 1/r_{g} = m^{2}_{P}/2W$ and $\lambda_{min}$, associated with the particle of mass m, is of the order of the gravitational radius ($m_{p}$ is the Planck mass, $10^{-5}$ g). Solving the integrals from Eq. (3.3) and (3.4), one obtains
\begin{equation}
\epsilon = \frac{m^{4}}{32 \pi^{2}} \left[x \sqrt{1+x^{2}}+2x^{3} \sqrt{1+x^{2}}-\ln(x+\sqrt{1+x^{2}})\right]
\label{3.5}
\end{equation}
and
\begin{equation}
\sigma = \frac{m^4}{32 \pi^2} \left[-x \sqrt{1+x^2}+\frac{2}{3} x^3 \sqrt{1+x^2}+\ln(x+\sqrt{1+x^2})\right],
\label{3.6}
\end{equation}
where $x = m^{2}_{P}/2mW$. With V fixed, we could find W from \eqref{3.5}. However, it is impossible to get an analitical expression because of the logarithm in the r.h.s. Therefore, we look for some approximate solution.
\\1). $x << 1$ ; $W >> m^{2}_{P}/2m \approx 10^{16} g$.
\\As we shall see, this case corresponds to the macroscopic situation. We have
\begin{equation}
\sqrt{1+x^2} \approx 1+\frac{x^2}{2}-\frac{x^4}{8}
\label{3.7}
\end{equation}
and
\begin{equation}
\ln(1+x+\frac{x^2}{2}) \approx x-\frac{x^3}{6}+\frac{x^4}{4}-\frac{x^5}{20}.
\label{3.8}
\end{equation}
Keeping terms up to the 5-th order, \eqref{3.5} yields now
\begin{equation}
\epsilon = \frac{m^4}{12 \pi^2} x^3 \left(1-\frac{3x}{32}+\frac{111 x^2}{320}\right).
\label{3.9}
\end{equation}
Negleting the last two terms in the paranthesis, one get
\begin{equation}
W^{4} = \frac{m^{6}_{P}}{96 \pi^2} m V
\label{3.10}
\end{equation}
(written in full, \eqref{3.10} appears as $W^{4} = (c^{14}/96 \pi^2 G^{3}) mV$). It is worth to note that the energy W does not depend on the Planck constant but only on c and V. It is proportional to $V^{1/4}$ and, therefore, $\epsilon$ is not constant.It decreases with V. In other words, the energy density is a function of the initial volume.
\\Another form of the expression $(3.10)$ would be
\begin{equation}
\left(\frac{W}{W_{P}}\right)^{4} = \frac{1}{96 \pi^{2}} \frac{m}{m_{P}} \left(\frac{L}{L_{P}}\right)^{3}
\label{3.11}
\end{equation}
where $L_{P}$ is the Planck length.
\\Let us find now for what values of L, \eqref{3.11} is valid. We must have $W >> m^{2}_{P}/2m$ or $V >> 6 \pi^{2} m^{2}_{P}/m^{5}$. Taking $V \approx L^{3}$, we obtain $L >> 1~mm$. It means the approximation $x << 1$ corresponds to linear dimensions of the chosen vacuum region much larger than $1~mm$.
\\From \eqref{3.11}, with, for example, $L \approx 10^{5}~ cm$, we obtain $W \approx 10^{39}$~ erg and $\epsilon \approx 10^{24}~ erg/cm^{3}$. The energy is, of course, very large but it is better than infinite.
\\Summing up the \eqref{3.5} and \eqref{3.6}, one obtains
\begin{equation}
\epsilon + \sigma = \frac{m^{4}}{12 \pi^{2}} x^{3} \sqrt{1+x^{2}} ,
\label{3.12}
\end{equation}
which could be written as 
\begin{equation}
\frac{\epsilon + \sigma}{\epsilon_{m}} = \frac{1}{12 \pi^{2}} x^{3} \sqrt{1+x^{2}},
\label{3.13}
\end{equation}
With $x<<1$ or $\epsilon + \sigma << \epsilon_{m}$, where $\epsilon_{m} = m^{4}$ is the energy density associated with the mass m, one has $\epsilon + \sigma \approx 0$. Put it differently, in this case the energy - momentum tensor of the quantum fluctuations of a massive scalar field is of $\Lambda$ - type \cite{YZ}, $\Lambda$ being the ``cosmological constant''
\begin{equation}
\Lambda = 8 \pi \epsilon .
\label{3.14}
\end{equation}
With $\epsilon$ from the previous example, we have $\Lambda \approx 10^{-48} \epsilon = 10^{-24} cm^{-2}$ and the radius of curvature associated to it is $R_{c} = \Lambda^{-1/2} = 10^{12} cm$. Since $R_{c} >> L$, valid for any $L >> L_{P}$, the de Sitter spacetime is practically flat inside the region considered. We know that there is a time - dependent ``comoving'' coordinate system where the spatial part of the metric is expanding. Hence, the quantum zero point fluctuations lead to a local vacuum expansion.
\\ 2).~$x >> 1$ ; $W << m^{2}_{P}/2m$. 
\\We have now $\sqrt{1+x^{2}} \approx x$ and \eqref{3.5} appears as 
\begin{equation}
\frac{W}{V} = \frac{m^{4} x^{4}}{32 \pi^{2}} \left(2+\frac{1}{x^{2}}-\frac{ln2x}{x^{4}}\right)
\label{3.15}
\end{equation}
A new approximation leads to 
\begin{equation}
\frac{W}{V} = \frac{m^{4} x^{4}}{16 \pi^{2}}
\label{3.16}
\end{equation}
or, using the expression for x
\begin{equation}
W^{5} = \frac{m^{8}_{P}}{256 \pi^{2}} V.
\label{3.17}
\end{equation}
With all fundamental constant, $W^{5} = (\hbar c^{17} / 256 \pi^{2}G^{4})~V$.
\\It is clear that the zero point energy in this case does not depend on the mass m of the field. In addition, the same approximation gives $\sigma = \epsilon/3$, the equation of state for radiation. Therefore, the Lorentz invariance of the vacuum is preserved as the scalar field becomes massless in the approximation used (with $x << 1$, the Lorentz invariance was assured by $\epsilon = -\sigma$). As we already noticed, we took here $\epsilon = W/V$ and not $dW/dV$, sice $W(V)$ is a power function and the difference is negligible.
\\Let us see now for what range of L we have $W << m^{2}_{P}/2m$, the condition of validity of (3.17). The restriction 
\begin{equation}
\frac{m^{8}_{P}}{256 \pi^{2}} V << \frac{m^{10}_{P}}{32 m^{5}}
\label{3.18}
\end{equation}
must be fulfilled.
\\A simple calculation gives $L << 1 mm$, the same limit as for the case 1.
\\A more useful form of (3.17) would be 
\begin{equation}
\frac{W}{W_{P}} = \left(\frac{1}{16 \pi}\right)^{2/5} \left(\frac{L}{L_{P}}\right)^{3/5}.
\label{3.19}
\end{equation}
With, for instance, $L \approx 10^{-8}~cm$, $W \approx 10^{30}~erg$ and $\epsilon \approx 10^{54}~erg/cm^{3}$. A comparison with \eqref{2.6} for the massless field shows that the functions $W(V)$ are similar, excepting the numerical factor and the range of validity.
\\~3).~$x = 1$~;~ $W = m^{2}_{P}/2m$.
\\Putting $x = 1$ in \eqref{3.5}, one obtains
\begin{equation}
\frac{W}{V} \approx \frac{3m^{4}}{32 \pi^{2}}.
\label{3.20}
\end{equation}
Combining with the condition $W = m^{2}_{P}/2m$, we have
\begin{equation}
\frac{V}{V_{P}} = 50 \left(\frac{m_{P}}{m}\right)^{5},
\label{3.21}
\end{equation}
whence $L = V^{1/3} \approx 1~mm$, which is the boundary between the previous two regions. This ``critical'' length may be related to the fundamental scale of quantum gravity \cite{KM} in the brane world scenarios. For $L << 1~mm$ the equation of state is $\sigma = \epsilon/3$ as for dark matter, while for $L >> 1~mm$ we have $\sigma = -\epsilon$, as for dark energy.

\section{CONCLUSIONS}
We have proposed in this paper a model to render finite the zero point energy of quantum fields. What we consider to be new with respect to the previous attempts is the appearance of Newton's constant, i.e. gravity, to make W finite.
\\The ``transition'' length $L \approx 1~mm$ might be related to the brane world models where the Newton law is supposed to undergo changes for L below 1 mm. With $L >> 1~mm$ (the macroscopic situation), W has a ``classical'' expression (no $\hbar$ appears in its formula) and, in addition, the equation of state $\sigma = -\epsilon$ leads to a Lorentz invariant vacuum, giving rise to a nonzero ``cosmological constant''. Anyway, the radius of curvature of the de Sitter spacetime is much larger than the linear dimension L of the domain considered and therefore the spacetime is practically flat.
\\We found that the energy density $\epsilon$ depends upon the volume of the region taken into consideration, decreasing with V.
\\An open question would be the approximation used for evaluating W for the two ranges of L. With a higher accuracy we could improve the results. Another open problem consists in the measurement of W, reminding the fact that its formula is valid for time intervals shorter than L.


\begin{thebibliography}{8}
\bibitem {DS}
D.W. Sciama, ``Black holes and fluctuations of quantum particles : An Einstein syntesis'', in Relativity, Quanta and Cosmology in the development of the scientific thought of Albert Einstein, vol. II, New York (1979).
\bibitem {LF}
L.H. Ford, ``Quantum vacuum energy in closed universe'', Phys.Rev.D14, 3304 (1976).
\bibitem {HH}
D.K.Hong and S.D.H. Hsu, ``Brane world confronts holography'', Prepriny hep-th/0401060 (2004).
\bibitem {RB}
R.Bousso, ``The holographic principle'', Preprint hep-th/0203101 (2002).
\bibitem {AA}
A.Aste, ``Holographic entropy bound from gravitational Fock space truncation'', Preprint hep-th/0409046 (2004).
\bibitem {TP}
T.Padmanabhan, ``Dark energy : the cosmological challenge of the Millennium'', Preprint astro-ph/0411044 (2004).
\bibitem {WS}
D.W.Sciama, ``Quantum field theory, horizons and thermodynamics'', Adv.Phys.30, 327 (1981).
\bibitem {YZ}
Ya.B.Zeldovich, ``Cosmologhiceskaia postaiannaia i teoriia elementarnih ceastitz'', Usp.Phys.Nauk 95, 209 (1968).
\bibitem {KM}
K.A.Milton, ``Dark energy as evidence for extra dimensions'', Preprint hep-th/0210170 (2002).
\end{thebibliography}
\end{document}